\documentclass[twocolumn,aps,prb,floatfix,showpacs]{revtex4}
\usepackage{graphicx}
\begin{document}

\title{Evidence of strong correlations at the van Hove
  singularity in the scanning-tunneling spectra of superconducting
  Bi$_{2}$Sr$_{2}$CaCu$_{2}$O$_{8+\delta}$ single crystals}

\author{Jouko Nieminen}
\email{jouko.nieminen@tut.fi}
\affiliation{Department of
Physics, Tampere University of Technology, P.O. Box 692, FIN-33101
Tampere, Finland~}
\affiliation{Department of Physics, Northeastern
University, Boston~}

\author{Ilpo Suominen}
\affiliation{Department of
Physics, Tampere University of Technology, P.O. Box 692, FIN-33101
Tampere, Finland}

\author{Tanmoy Das}\author{R.S. Markiewicz} \author{A. Bansil}
\affiliation{Department of Physics, Northeastern University, Boston}

\date{Version of \today}

\begin{abstract}

  We present realistic multiband calculations of scanning tunneling
  spectra in Bi$_2$Sr$_2$CaCu$_2$O$_{8+\delta}$ over a wide doping
  range. Our modeling incorporates effects of a competing pseudogap
  and pairing gap as well as effects of strong electronic
  correlations, which are included by introducing self-energy
  corrections in the one-particle propagators. The calculations
  provide a good description of the two-gap features seen in
  experiments at low energies and the evolution of the Van Hove
  singularity (VHS) with doping, and suggest a possible quantum
  critical point near the point where the VHS crosses the Fermi level.

\end{abstract}

\date{Version of \today}

\maketitle

\section{Introduction}

A curious, topological feature of any two dimensional electronic band
is a saddle-point Van Hove singularity (VHS), which can lead to a
crossover from electron-like to hole like Fermi surfaces with doping,
accompanied by a logarithmically diverging density-of-states (DOS) and
a vanishing Fermi velocity\cite{VHS0}.  While the divergence itself
can easily be smoothened by disorder or interlayer coupling, the
presence of a peak in the DOS is a robust feature of the underlying
spectrum. In hole-doped cuprates, most theoretical studies of doping
dependence find enhanced tendencies for superconductivity or competing
phases -- including the possibility of phase separation -- when this
VHS approaches the Fermi level $E_F$.\cite{VHS1,Jarrell} This scenario
seems to work for La$_{2-x}$Sr$_x$CuO$_4$, where the VHS nearly
coincides with $E_F$ at optimal doping\cite{VHS4}, but in most other
cuprates, with a larger second-neighbor hopping parameter $t'$,
optimal doping and the VHS doping seem to separate further as $T_c$
increases.\cite{KamRa} Notably, early angle-resolved photoemission
spectroscopy (ARPES) experiments found evidence for a peak-dip-hump
structure, where an electronic mode pushed a narrow coherent band with
a prominent VHS (`peak') very close to the Fermi level of most
cuprates, while pushing incoherent spectral weight away from $E_F$
(`hump'), thereby leaving a `dip' in the spectrum near the mode
energy.\cite{pdh} However, the coherent VHS peak has proved elusive in
these materials, and there is controversy as to whether it can be seen
in tunneling spectroscopy\cite{VHS2,VHS3}.

Here we explore this issue by comparing recent scanning tunneling
microscopy and spectroscopy (STM/STS) data with a realistic multiband
modeling of the spectra in Bi$_2$Sr$_2$CaCu$_2$O$_{8+\delta}$
(Bi-2212). The model is able to reproduce the STM features in detail, 
shedding light on the role of the VHS and the ordering phenomena 
involved in `two gap' physics. 

Our doping-dependent tunneling spectrum is calculated using a
multiband Nambu-Gorkov Green's function tensor formalism.  The main
technical innovation of this work is to incorporate a variety of
intermediate coupling effects into the tunneling formalism via a
tensor self-energy extracted from quasiparticle-GW (QP-GW)
calculations.\cite{tanmoy2gap}.  We model the two competing order gaps
as due to BCS-type SC coupling\cite{NLMB,nieminenPRB} and
antiferromagnetic order.  To these we add intermediate coupling
self-energies, calculated in a GW model.  Our QP-GW
self-energies have been shown to capture a wide range of phenomena in
cuprates over an extended range of doping and excitation energies,
including the relationship between magnetic order and superconductivity,
and the so-called waterfall physics.\cite{basak} Notably, for a
realistic consideration of the features in STM/STS, it is essential to
utilize a multiband framework. This makes it possible to incorporate
filtering and tunneling path effects due to the BiO and SrO layers
separating the vacuum and the cuprate layers, and the important role
of the apical oxygen and Cu $d_{z^{2}}$ orbitals.\cite{NLMB,nieminenPRB}

While the one band QP-GW model employed to obtain the self-energies used
in this study is nearly parameter-free\cite{footB4}, the present
calculations unavoidably introduce additional parameters due to two
separate factors.  Firstly, the pairing gaps are treated as
parameters, since the Eliashberg theory cannot accurately predict the
gap size, particularly in the pseudogap phase.  A second set of
parameters arises because STM experiments on the cuprates find that
the gap size varies in different local patches across a given
crystal. In our modeling, we assume that different patches represent
regions of different local
doping\cite{ZDW,Wise2,footB10,McElroy,Nunner}.  A strong inference
from our study is support for this local doping model since a number
of observed spectral features can be correlated with theory over a
wide range of energies and dopings.

This article is organized as follows. Section II describes our model
of electronic spectrum and gives a brief overview of the formalism of
tunneling calculations. Section III presents theoretical results for
the normal and the superconducting state as a function of doping and
compares and contrasts these results with the corresponding
experimental results, including issues related to the two-gap physics
and signatures of the VHS. Section IV discusses broader implications
of our study. This is followed by a summary of our conclusions in
Section V. Appendix A gives details of the interaction terms used in
the Hamiltonian, while Appendix B discusses details of the self-energy
corrections used for incorporating electron correlation effects in our
tunneling computations.  Appendix C summarizes a related model which
displays peak-dip-hump physics.

\section{Methodology}
\subsection{Green's function formulation of tunneling current}

Our computations are based on a multiband tight-binding model using
$4\times 116$ orbitals per unit cell, i.e., a fourfold basis taking
into account electrons and holes for both spin
directions.\cite{NLMB,nieminenPRB} For treating the magnetic order and
superconducting pairing, the tensor (Nambu-Gorkov) Green's function
${\cal G}$ is employed with the corresponding Dyson's
equation:\cite{Fetter}
\begin{equation}
  {\cal G} =  {\cal G}^0 +  {\cal G V G}^0,
\label{dyson1}
\end{equation}
where
\begin{displaymath}
  {\cal G} =
\left(
   \begin{array}{cccc}
G_{e \uparrow}& 0& 0&  F_{\uparrow \downarrow}\\
0&G_{e \downarrow}&  F_{\downarrow \uparrow} & 0\\
0& F^{\dagger}_{\downarrow \uparrow}& G_{h \uparrow}&0\\
F^{\dagger}_{\uparrow \downarrow}& 0 &0&  G_{h \downarrow}
   \end{array}
\right)~\textrm{with}~
{\boldmath c}_{\alpha} =
\left(
  \begin{array}{c}
c_{\alpha \uparrow}\\
c_{\alpha \downarrow}\\
c^{\dagger}_{\alpha \uparrow}\\
c^{\dagger}_{\alpha \downarrow}
  \end{array}
\right)
\end{displaymath}
where $G_{e}$ and $G_{h},$ denote the Green's functions for the
electrons and holes, respectively\cite{foot11}, and the matrix
elements of operator ${\cal V}$ represent the interaction terms of
Hamiltonian of Eq. (7), below.

In the tunneling calculations the filtering effect of the surface layers
is taken into account by using the Todorov-Pendry approach
\cite{Todorov} in which the
differential conductance $\sigma$ between orbitals of the tip ($t,t'$)
and the sample ($s,s'$) is given by\cite{NLMB,nieminenPRB}
\begin{equation}
\sigma = \frac{dI}{dV} = \frac{2 \pi e^2 }{ \hbar} \sum_{t t' s s'}
\rho_{tt'}(E_F)V_{t's} \rho_{ss'}^{}(E_F+eV)V_{s't}^{\dagger},
\label{conductance}
\end{equation}
where the density matrix
\begin{equation}
\rho_{s s'} = -\frac{1}{\pi}Im[G_{s s'}^{+}]
 = \frac{1}{2\pi i} \left( G^{-}_{s s'}
- G^{+}_ {s s'} \right) ,
\label{spectralfunctiona}
\end{equation}
is expressed in terms of the retarded and advanced electron Green function or
propagators. Eq.  \ref{conductance} differs from the
more commonly used Tersoff-Hamann formulation\cite{Tersoff} in that it
takes into account the details of the symmetry of the tip orbitals and
how these orbitals couple with the orbitals of the cuprate layer through
the filtering BiO and SrO layers.

Equation~\ref{spectralfunctiona} can be rewritten as\cite{nieminenPRB}
\begin{equation}
\sigma =
 \frac{2 \pi e^2 }{\hbar}\sum_{tt' c c'}
 \rho_{tt'}(E_F)M_{t'c}\rho_{cc'}(E_F+eV)M_{c't}^{\dagger},
\label{cupraldos}
\end{equation}
where
\begin{equation}
M_{tc} = V_{ts}G^{0+}_{sf}V_{fc},
\label{filter}
\end{equation}
gives the filtering amplitude between the cuprate layer and the
tip, and constitutes a multiband generalization of the filtering function
of Ref. \onlinecite{Balatsky}.  Note that the matrix element of the
density of states operator $\rho_{cc'}$ within the cuprate plane can
be recovered in terms of the spectral function:
\begin{equation}
\rho_{c c'}= -\frac{1}{\pi} \sum_{\alpha}
(G^{+}_{c\alpha}\Sigma{''}_{\alpha}G^{-}_{\alpha
   c'}+F^{+}_{c\alpha}\Sigma{''}_{\alpha}F^{-}_{\alpha c'})
\label{singlespec}
\end{equation}
Notably, the matrix $M$ itself is rather structureless, but it is very
selective in the way it couples the insulating layers and the cuprate
layer due to the term $V_{fc}$.\cite{nieminenPRB} In fact, for
symmetry reasons the tunneling channel passes through the $d_{z^{2}}$
orbital and not the $d_{x^{2}-y^{2}}$ orbital of the Cu atoms.

\subsection{Interaction Effects}

In order to meaningfully describe the doping dependence of the
tunneling spectra of the cuprates, various key interactions must be
included in the Hamiltonian. To this end, we introduce interactions
via a Hubbard $U$ on the Cu sites, which are approximated by a mean-field
antiferromagnetic order plus GW-type self-energy associated with spin
and charge fluctuations.  We model the two competing order gaps as
being due to a BCS-type SC coupling, and an antiferromagnetic
order. Superconductivity is included by adding a d-wave pairing
interaction term $\Delta$ and the AFM order\cite{footB9} is included
by introducing a Hubbard term $Um$, leading to the Hamiltonian:

\begin{equation}
  \begin{array}{c}
\hat{H}= \sum_{\alpha\beta\sigma}
(\varepsilon_{\alpha}c^{\dagger}_{\alpha \sigma} c_{\alpha \sigma}+
V_{\alpha \beta}
c^{\dagger}_{\alpha \sigma} c_{\beta\sigma}+
\Delta_{\alpha \beta} c^{\dagger}_{\alpha \sigma}
c^{\dagger}_{\beta -\sigma} + \\
\Delta_{\beta \alpha}^{\dagger}
 c_{\beta -\sigma} c_{\alpha \sigma} )
- 
\sum_{\alpha} U m_{\alpha}(c^{\dagger}_{\alpha \uparrow} c_{\alpha \uparrow}
-c^{\dagger}_{\alpha \downarrow} c_{\alpha \downarrow})
  \end{array}
\label{hamiltonian}
\end{equation}
with the real-space creation (annihilation) operators
$c^{\dagger}_{\alpha \sigma}$ ($c_{\alpha \sigma}$). Here $\alpha$ is
a composite index for the type of orbital and its site, and $\sigma$
is the spin index.  $\varepsilon_\alpha$ is the on-site orbital
energy, while $V_{\alpha\beta}$ is the hopping integral between
orbitals $\alpha$ and $\beta.$\cite{abfoot1,new1,new2} The band
structure related parameters $\varepsilon_{\alpha}$ and $V_{\alpha
  \beta}$ are the same as those used in Ref. \onlinecite{nieminenPRB}.
Details of the interaction parameters $U$, $m$, and $\Delta$ are given
in Appendix A.

In order to model effects of the low and high energy bosonic couplings
and background scattering in the electronic system, we add a
self-energy correction in the propagators, which consists of three
distinct contributions:
\begin{equation}
  \label{eq:totalself}
  \Sigma_{\alpha \beta}(\varepsilon) = \Sigma^{L}_{\alpha \beta}(\varepsilon)+
 \Sigma^{H}_{\alpha \beta}(\varepsilon)+
\Sigma^{imp}_{\alpha \beta}(\varepsilon).
\end{equation}
Here, superscripts $L$ and $H$ refer to corrections due to the low-
and high-energy bosonic couplings, respectively, while the superscript
$imp$ refers to the correction term arising from background impurity
scattering.  For $\Sigma^L$ we use a phonon contribution based on a
Debye spectrum and $\Sigma^H$ is obtained within the framework of a
self-consistent GW scheme. Details of the self-energy corrections of
Eq. 8, including an overview of the GW-scheme used, are given in
Appendix B.  We note that in the theoretical STS spectra discussed in
Section III below, the electronic contribution $\Sigma^H$ dominates in
producing key features in the spectra, while $\Sigma^{imp}$ yields a
realistic broadening of the spectra and $\Sigma^L$ contributes to the
sharpness of the coherence peaks.

\begin{figure}
  \includegraphics[width=0.4\textwidth,angle=0]{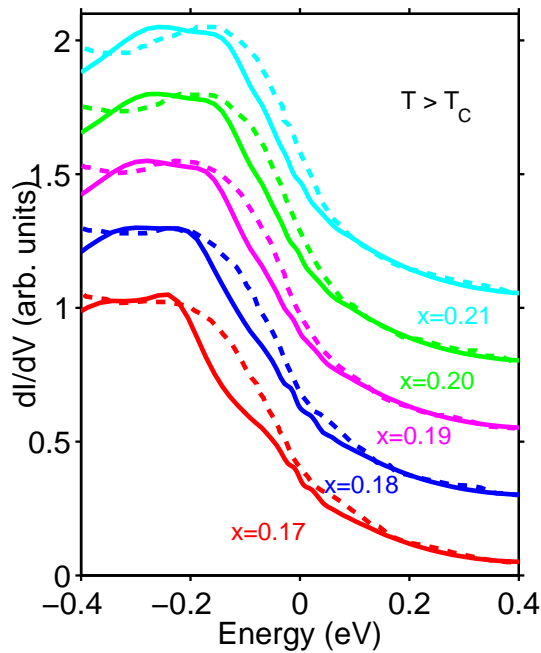}
  \caption{ (Color online) Comparison of the theoretical (solid)
    and experimental (dashed, Ref. \onlinecite{Yazdani}) normal state
    STS spectra over the hole doping range $x=0.21-0.17$.}
\label{normal_0}
\end{figure}

\section{Results}
\subsection{Normal State Spectra}

Fig. 1 compares the normal state spectra taken above the
superconducting transition temperature $T_c$ where superconductivity
is absent although a pseudogap could be present.  Experimental data
over a wide energy range are available only near and above optimal
doping where the pseudogap is small or zero.\cite{Yazdani} In
Ref. \onlinecite{Yazdani}, the superconducting gap map of an
inhomogenous sample was first measured at low temperature to determine
the local gap $\Delta$ in different domains on the surface.  The
temperature was then raised above the highest superconducting
transition temperature, and the corresponding normal state spectra
were measured.  Along these lines\cite{ZDW,Wise2}, we use the measured
low-$T$ gap values varying from 20 meV to 36 meV to adduce the local
doping, which yields doping values varying over $x=0.21 - 0.17$ as
shown in Fig. 1. The theoretical as well as the experimental spectra
in Fig. 1 are seen to be quite featureless, except for the pronounced
asymmetry between high positive and negative bias voltages, which we
have analyzed previously.\cite{NLMB} The most prominent feature in the
spectra is a large hump-like feature located around -200~meV binding
energy for $x$= 0.21, which shifts away from the Fermi level to higher
binding energies and becomes smoother as the doping is decreased. Our
calculations are seen to reproduce this hump-like feature and its
characteristic shift and smoothening with doping in substantial accord
with experimental results. It should be noted that over the doping
range considered in Fig. 1, pseudogap effects in our computations are
quite small. We have verified this by carrying out a series of
computations where the pseudogap was artificially turned off (not
shown in Fig. 1 for brevity), suggesting that the pseudogaps have
closed at the temperature of 93K at which the experimental data are
taken.

\begin{figure}
  \includegraphics[width=0.40\textwidth,angle=0]{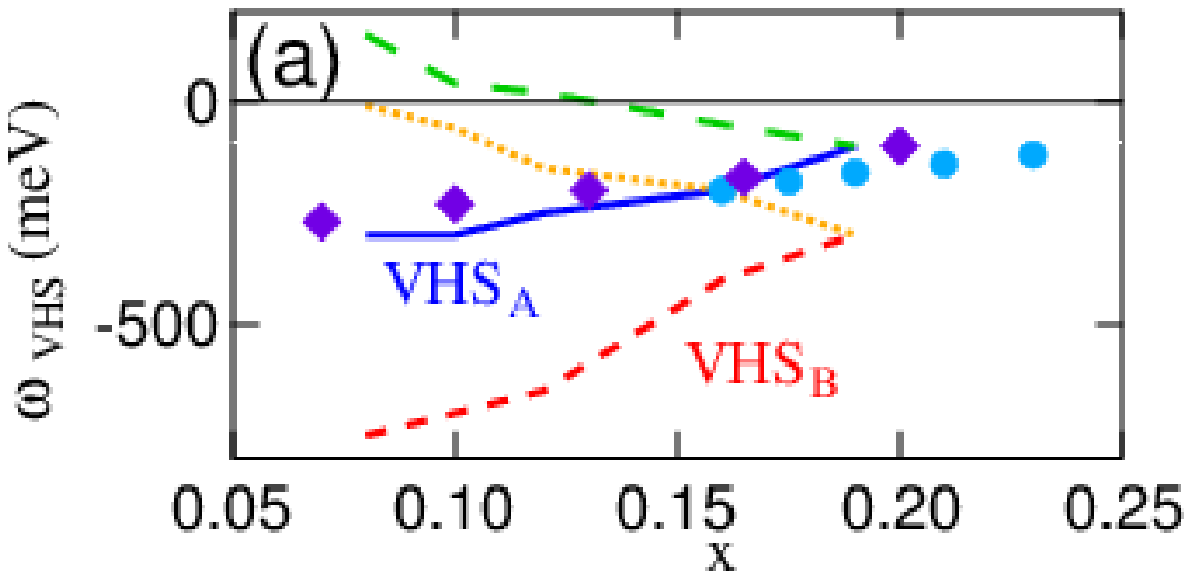}
  \includegraphics[width=0.40\textwidth,angle=0]{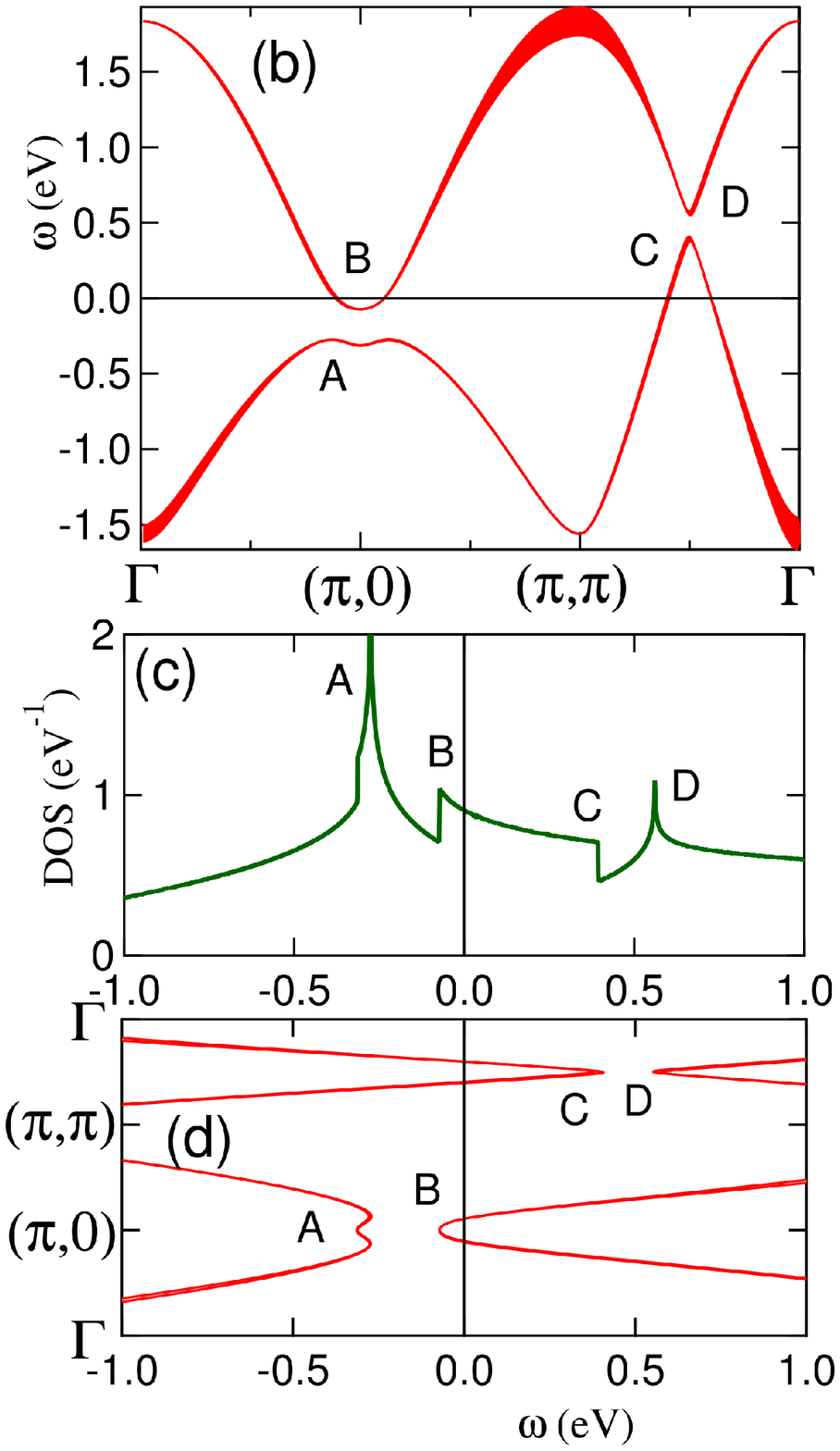}
 \caption{ (Color online) (a) Derived AF state phase diagram. The
   filled blue circles [Ref.~\onlinecite{Yazdani}] and violet diamonds
   [Ref.~\onlinecite{McElroy}] are the hump features for Bi2212
   derived from recent STM measurements.  These are compared to a
   calculated doping dependence of the bonding [B] and antibonding [A]
   VHS (red short-dashed and blue solid lines, respectively) in
   Bi2212, based on the present analysis. (b) Calculated dispersion at
   $x=0.15$, with feature A, the VHS of the lower magnetic band,
   corresponding to the blue solid line [or red short-dashed line, for
   the bonding band] in (a), and feature B, the bottom of the upper
   magnetic band, corresponding to the green long-dashed line [orange
   dotted line] in (a). The thickness of the lines represents the
   spectral weight due to the AF structure factor.  (c) Corresponding
   DOS, showing features derived from A, B, C, and D in (b).  (d)
   Blow-up of the near-$E_F$ dispersion.}
\label{phasediagN}
\end{figure}

Figure~\ref{phasediagN}(a) gives a plot of the energies of various
features in the theoretical and experimental STS spectra of Fig. 1 to
allow a more quantitative discussion of the doping dependencies of
these features and their correlation with the phase diagram of
Bi2212. The model includes bilayer splitting, and the solid blue
[short-dashed red] curve represents the VHS peak of the antibonding
[bonding] band.  We see that the experimental normal state hump (blue
filled circles)\cite{Yazdani} is in good agreement with the doping
dependence of the antibonding VHS.  The calculated VHS position does
not change significantly in the superconducting state, so we also plot
the experimental hump feature from the superconducting state (violet
filled diamonds)\cite{McElroy}, from Fig.~3 below.  When an AF
pseudogap turns on, a second feature appears [green long-dashed and
orange dotted lines for the antibonding and bonding bands
respectively].  Frames (b) and (c) explain the origin of this feature
for the antibonding band, labeled $B$.  
These figures are based on a
three-band model, containing a Cu $d_{x^2-y^2}$ and two O $p$
orbitals, in which bilayer splitting is neglected, to clarify the
origin of the features.  From Fig.~3(b), it can be seen that the AF
gap has a strong $k$-dependence, splitting the band into upper and
lower magnetic bands (U/LMBs).  Features $A$ and $D$ are the VHSs of
the LMB and UMB respectively, while feature $B$ is the bottom of the
UMB and feature $C$ the top of the LMB.  Note the characteristic form
of the DOS associated with each feature in Fig.~2(c).

As the AF pseudogap shrinks to zero, features $A$ and $B$ [and $C$ and
$D$] merge.  That is why there is no near-$E_F$ feature in the three
spectra in Fig.~1 with higher dopings.  For lower doping a feature
corresponding to $B$ appears in the low-T theoretical curves, but
seems to be absent in the $T$=93K experimental data of Fig.~1,
suggesting a $T$-dependent pseudogap closing.  Note that at lower
doping feature $B$ moves closer to the Fermi level, and should persist
to higher temperatures, as the pseudogap is larger.

\subsection{Superconducting State spectra}

\begin{figure}
  \includegraphics[width=0.5\textwidth,angle=0]{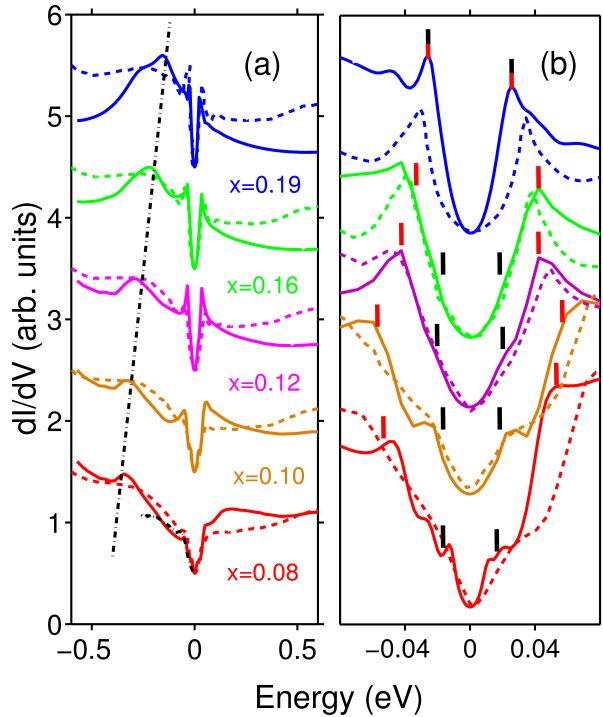}
  \caption{ (Color online) (a) Theoretical (solid) and experimental
    (dashed, Ref. \onlinecite{McElroy}) superconducting state spectra
    over the hole doping range from $x=0.19$ at the top to $x=0.08$ at
    the bottom of the figure.\cite{footB1} For comparison, the
    experimental ARPES spectrum\cite{ARPES} at $x=0.08$ (black dashed
    curve) is shown. The dot-dashed line is drawn through the hump
    feature in the computed spectra as a to guide the eye. (b) Blowup of
    theoretical spectra near the Fermi level is compared with the
    corresponding data of Ref.~\onlinecite{Lawler}. The black and red
bars indicate the gap widths shown as empty and filled diamonds in Fig. 4. }
\label{phasediag0}
\end{figure}

Figure~ \ref{phasediag0}(a) compares our theoretical
results\cite{footB1} with the corresponding experimental
spectra\cite{McElroy,Lawler} at five representative values of hole
doping, with doping decreasing from the optimally doped SC phase at
$x=0.19$ at the top to the underdoped pseudogap phase for $x=0.08$ at
the bottom of the figure. With reference to Fig. 1, we see that the
hump-like feature at negative energies and its characteristic doping
dependency in the normal state is present also in the superconducting
state spectra of Fig.~\ref{phasediag0} (plotted in
Fig.~\ref{phasediagN}(a)).  The main difference between the normal and
superconducting state spectra is the opening up of a complex gap
structure at low energies, which is highlighted in the expanded view
of Fig.~\ref{phasediag0}(b).\cite{footB2} We note here with reference
to the lowest set of curves in Fig.~3(a) as an example that the STM
spectra resemble the antinodal ARPES spectra.\cite{ARPES} Moreover,
the angle-dependence of the gap function adduced from the STM
quasiparticle interference maps is in good agreement with the gaps
obtained via ARPES measurements.\cite{McE} These results support our
use of the local gap to determine the local doping of the patches in
STM data.

\begin{figure}
  \includegraphics[width=0.5\textwidth,angle=0]{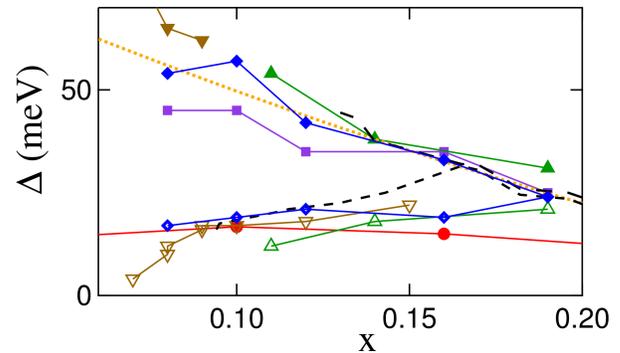}
  \caption{ (Color online) Two gap physics in the superconducting
    state.  The two theoretical gaps from Fig.~3(b) are plotted as
    filled and empty blue diamonds and compared to experimental ARPES
    results from Ref.~\onlinecite{Lee} (green triangles) and
    Ref.~\onlinecite{Tanaka} (brown inverted triangles) and
    corresponding Raman results for HgBa$_2$CuO$_{4+\delta}$ (Hg1201)
    (black long- and short-dashed lines) from
    Ref.~\onlinecite{Tacon2}.  Shown also is the input SC gap
    parameter, violet squares, the expected position of a VHS, based
    on an experimental ARPES dispersion\cite{MikeN}, orange dotted
    line, and a calculated SC gap\cite{Bob_sc}, red circles, to
    illustrate the doping dependence expected for a spin-pairing
    mechanism.  }
\label{phasediagSC}
\end{figure}

Figure~\ref{phasediagSC} gives a plot of the two low-energy gaps from
the theoretical STS spectra of Fig. 3, shown as filled and empty blue
diamonds.  These are seen to be in good agreement with the `two gap'
features extracted from ARPES\cite{Lee,Tanaka} and with Raman
results\cite{Tacon,Tacon2} for HgBa$_2$CuO$_{4+\delta}$ (Hg1201).  The
remaining curves\cite{MikeN,Bob_sc} will be discussed below.
Interestingly, our computations reproduce the remarkable {\it
anticorrelation} displayed by the large hump feature at negative
energies in the experimental STS spectra in Figs. 2 and 3 in the way
it varies with doping in relation to the doping dependence of the size
of the gap\cite{Yazdani}. Specifically, with increasing doping, as the
hump moves closer to the Fermi energy and the density of states at the
Fermi energy increases, the size of the gap in Figs. 3 and 4 does not
become larger as we would expect from the BCS theory, but instead the
gap becomes smaller. The size of the coherence peaks, on the other
hand, is seen to grow with increasing doping in our theory in accord
with experiments.

\begin{figure} 
\includegraphics[width=0.5\textwidth,angle=0]{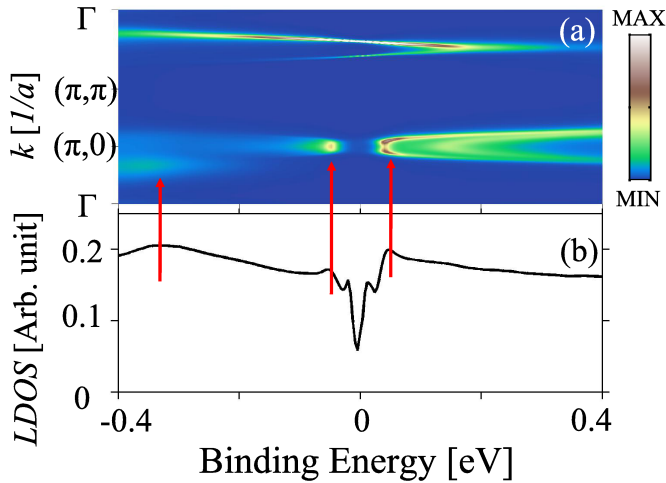}
\caption{(Color online) (a) Calculated $E-k$ spectral weight diagram
  for $x=0.08$ along high symmetry line
  $\Gamma-(\pi,\pi)-(\pi,0)-\Gamma$ in momentum space. White/brown
  color denotes high and blue color low intensity. (b) The
  corresponding total LDOS from Cu sites, with arrows relating three
  prominent LDOS peaks to spectral features in (a). Note that several
  peaks seen in LDOS in the low energy region do not show up in (a)
  because the associated spectral weight lies along the zone diagonal,
  which is not shown in (a) [see Fig. 6(b)].}
\label{ldos_1}
\end{figure}

Fig. 5 gives insight into the origin of the two gap physics in
Bi2212. For this purpose panel (a) gives a map of the spectral weight
obtained from the imaginary part of the one-particle Green's function
along the $\Gamma-X(\pi,0)-M(\pi,\pi)-\Gamma$ line in the
k-space. This map may be thought of as an effective theoretical E vs k
spectrum resulting after the self-energy correction $\Sigma$ has been
applied to the propagators. In the absence of self-energy corrections,
this map will consist of $\delta$-function peaks arising from the real
poles of the Green's function corresponding to the familiar energy
bands of Bi2212 modified by the the AFM and SC
orders.\cite{footB3,bansil12} At the doping of $x=0.08$ considered in
Fig. 5, the antibonding VHS lies around -300 meV, well below the Fermi
energy. After self-energy corrections are introduced, the VHS is
broadened substantially and yields the large hump-like structure in
the local density of states around -300 meV as seen from the LDOS plot
of Fig. 5(b) [leftmost arrow].  Due to bilayer
splitting\cite{footab,arpesab,MBRIXS,comptonab,positronab}, the
bonding VHS lies near -800~meV at this doping, and is not seen in
Fig.~5.

Next, we address the features near $E_F$ in Figs.~3(b) and 4 and how
these features relate to the two-gap physics, which has been discussed
in connection with Raman scattering\cite{Tacon,Tacon2}, angle-resolved
photoemission (ARPES)\cite{Lee,Tanaka}, and recent STM
studies\cite{Kohsaka08,pushp}.  Two different components of the gap
have been observed with different doping dependences\cite{Hufner}: a
nodal pairing gap $\Delta_n$ with a parabolic doping dependence
reminiscent of the superconducting $T_c$, and an antinodal gap
$\Delta^*$ which increases roughly linearly with underdoping, similar
to the pseudogap onset temperature $T^*$. The experimental near-gap
features\cite{Lawler} in Figs.~\ref{phasediag0}(b) and 4 clearly
reflect the presence of two gaps, which are well captured by the
present calculations. 
While the overdoped sample displays a single gap
feature, samples at lower doping show four features comprising two
gaps, an outer gap which grows with underdoping and an inner gap which
shrinks (see Fig. 4), in good agreement with the
two-gap scenario.  Coherence peaks are prominent in the overdoped
sample, but gradually disappear as doping is lowered, in good accord
with the computed STM spectra.  [In the undoped case we find a large
magnetic gap, with no evidence of superconductivity (not shown).]
From Fig.~3(b) it can be seen that our modeling reproduces the
electron-hole asymmetry of the experimental coherence peaks, which is
unexpected in the BCS formalism\cite{Hirsch}.

\begin{figure}
  \includegraphics[width=0.5\textwidth,angle=0]{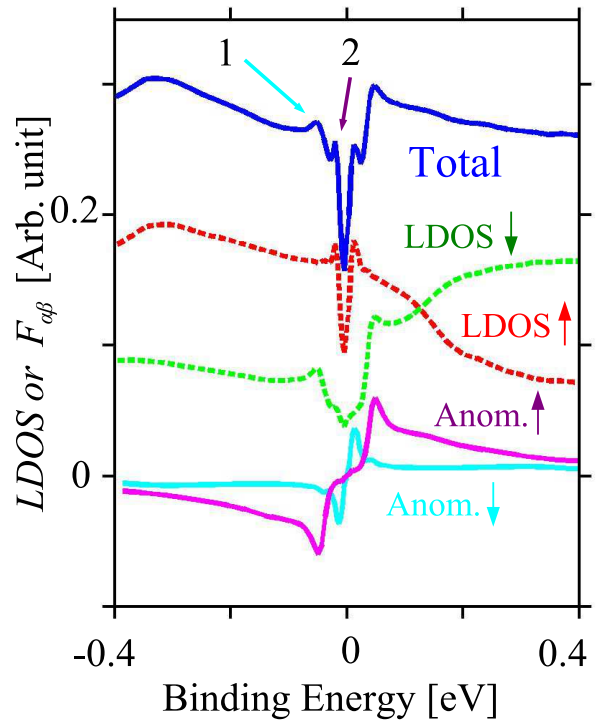}
  \caption{(Color online) Calculated spin resolved LDOS of the
    Cu-$d_{x2-y2}$ orbital for $x=0.08$: Total LDOS (blue solid line)
    and its decomposition into contributions from the lower magnetic
    band (LMB, red dashed line) and the upper magnetic band (UMB,
    green dashed line).  Also shown are the corresponding imaginary
    parts (cyan and magenta solid lines) of the anomalous matrix
    elements $F_{\alpha \beta}$ of the Nambu-Gorkov Green's function
    for the $d_{x2-y2}$ orbitals on two neighbouring Cu atoms. This
    matrix element indicates strong electron-hole hybridization or
    pairing effect at the pseudogap edge labeled by `1' and the in-gap
    peak labeled by `2' in the total LDOS (solid blue line). }
\label{ldos_2}
\end{figure}

Fig. 6 gives the decomposition of the LDOS of Fig. 5(b) for x=0.08
into the spin-resolved contributions from the upper and lower magnetic
bands (UMB and LMB). In our computations there is a small magnetic gap
separating the UMB and LMB, leading to electron pockets near $(\pi,0)$
and hole pockets near $(\frac{\pi}{2},\frac{\pi}{2})$. The two gaps in
the total LDOS curve in Fig. 6 are identified by the arrows labeled 
1 and 2, and are plotted in Fig.~\ref{phasediagSC} as blue diamonds.
In order to clarify the role of pairing in these two features,
Fig.~\ref{ldos_2} also shows the anomalous matrix elements
$F_{\alpha\beta}$ of the Nambu-Gorkov Green's function\cite{foot2},
which can be used to monitor the pairing effect. The contribution of
the anomalous elements to both the in-gap and gap-edge peaks indicates
that pairing is involved in both features. Notably, the gap features
in Fig.~3(b) more closely resemble the UMB features in Fig. 6,
indicating that the tunneling matrix element couples differently to
the LMB and UMB states.

\begin{figure}
  \includegraphics[width=0.5\textwidth,angle=0]{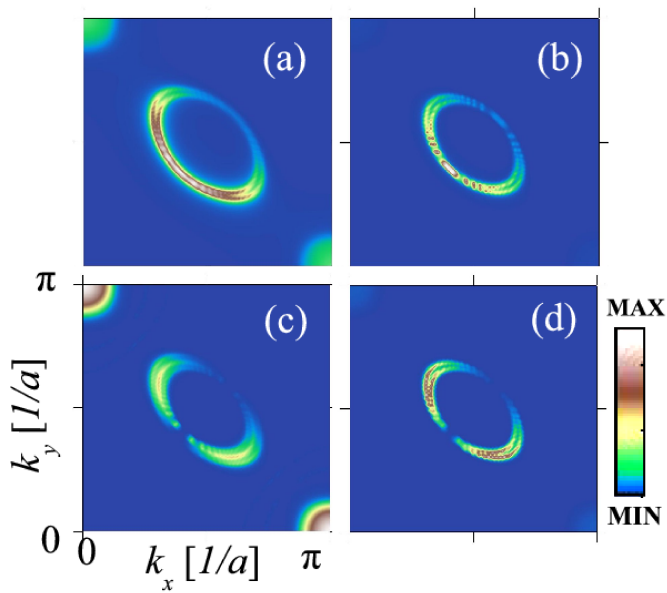}
  \caption{(Color online) (a) Spectral map at constant energy for
    $x=0.08$ for the regular matrix elements of the Nambu-Gorkov
    Green's function. The energy corresponds to the lower edge of the
    pseudogap, i.e. the energy of the peak marked 1 in Fig. 5. (b)
    Same as (a), except the energy corresponds to the energy of peak 2
    in Fig. 5. (c) Same as (a), except it gives the spectral map for
    anomalous matrix elements of the Green's function at the energy of
    peak 1. (d) Same as (c) for the peak 2. Arcs in (c) and (d) have
    zero weight along the nodal line, where the gap vanishes.}
\label{ldos_3}
\end{figure}

Further insight into pairing can be obtained by considering the
evolution of dispersions with binding energy.  Fig.~7 shows
spectral maps in momentum space at the energies of the coherence peaks
for the lower edge of the pseudogap in frame (a) and for the lower
in-gap peak in frame (b). The corresponding anomalous spectral weights
are shown in frames (c) and (d).  The peak at the pseudogap edge is
associated with the top of the $(\pi ,0)$ pocket as can be seen from
the strong pairing intensity in Fig.~7(c).  The in-gap peak is
associated with the pairing gap on the $(\pi /2,\pi /2)$ pocket,
Fig.~6(d), where AF order is absent.  Note that at this energy the
$(\pi ,0)$ pocket is absent, Fig.~7(b).  The $(\pi ,0)$ peak is found
to predominantly represent AF order with a Bilbro-McMillan-like
dressing by superconductivity\cite{bilbro}, while the $(\pi /2,\pi
/2)$ peak is a pure SC gap.  Note in Fig.~7(b) that the spectral
weight of the $(\pi /2,\pi /2)$ pocket resembles an arc with peak
intensity at the AF zone boundary as seen in STM studies. With
decreasing doping the gap between the LMB and UMB increases, leading
to a monotonically increasing pseudogap energy.  In contrast, the
Fermi surface on the LMB near $(\pi /2,\pi /2)$ shrinks monotonically
since its area is proportional to $x$.  For a $d$-wave gap which
vanishes at $(\pi /2,\pi /2)$, its maximum value must ultimately
decrease with underdoping, thereby explaining the opposite doping
dependence of the two gaps.

\subsection{Comparison with Bi2201}

\begin{figure}
  \includegraphics[width=0.5\textwidth,angle=0]{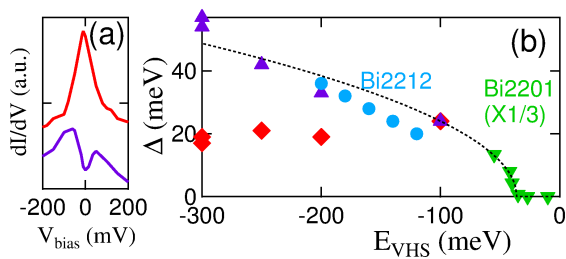}
  \caption{(Color online) (a) Tunneling $dI/dV$ spectra for Bi2201
    corresponding to $E_{VHS}$ = -10 meV (upper, red line) and -56~meV
    (lower, violet line, offset for clarity), after
    Ref.~\onlinecite{VHS3}.  (b) Plots of $\Delta$ vs $E_{VHS}$ for
    Bi2212 (red diamonds and violet triangles from
    Ref.~\onlinecite{Lawler} and blue circles from
    Ref.~\onlinecite{Yazdani}) and Bi2201 (inverted green triangles
    from Ref.~\onlinecite{VHS3}).  Dotted line indicates a square-root
    dependence of $\Delta$ on $E_{VHS}+36$~meV.  }
\label{Bi2201}
\end{figure}

To gain more insight into the relation of $E_{VHS}$ and the
superconducting gap $\Delta$, in Fig.~8 we compare the present results
with recent experiments on Bi$_2$Sr$_2$CuO$_{6+\delta}$
(Bi2201).\cite{VHS3} We see that the shape of the VHS at highest
doping is consistent with the calculated shape in Bi2212 at lower
doping 
compare the top curves in Fig.~3(a) and Fig.~8(a) 
but
that it gets considerably broadened for slightly lower doping, lower
curve in Fig.~8(a).  This may explain why our calculations in
Fig.~3(a) underestimate the VHS broadening, as the QP-GW technique
tends to produce too little broadening at higher energies.

The decrease of $\Delta$ with increasing $E_{VHS}$, Fig.~8(b), is
quite similar in both materials, although with somewhat faster change
in Bi2201 (note change of scale).  The remarkable resemblance to the
dotted curve is unexpected, and deserves some comment.  The dotted
curve represents the relation
\begin{equation}
  \label{eq:VHSeq}
  \Delta^2+\Delta_0^2 = A|E_{VHS}|,
\end{equation}
with $\Delta_0\sim 18~meV$, and $A\sim\Delta_0/2$.  The general form
$\sum_i\Delta_i^2=constant$ is a common finding for competing orders,
but in Eq.~\ref{eq:VHSeq} the meaning of $\Delta_0$ is unknown, and
the right-hand side is not constant, but vanishes when the VHS is at
the Fermi level.  This is suggestive of yet another quantum critical
point (QCP) in the cuprates, but this one associated with the
VHS.\cite{Jarrell,RM7} Unlike most VHS-related features, in this case
the gap scales to zero at the VHS, suggestive of nesting of the
antinodal parts of the Fermi surface.\cite{RM7}

Remarkably, QCPs have been proposed at three dopings in the cuprates:
in the underdoped regime, where superconductivity is lost in a
superconductor-insulator transition, near optimal doping, and now in
the overdoped regime, where superconductivity vanishes near a VHS.
Equation~\ref{eq:VHSeq}, or its generalization to multiple order
parameters, offers a rationale for such a series of transitions, as
the shrinking nesting inaugurates an ever more desparate competition
among phases.

\section{Discussion}

The present calculations strongly support the idea that a competing
order pseudogap provides a good description of two gap physics in hole
doped cuprates.\cite{foot5,tanmoy2gap} The model offers new insights
into the role of the VHS.  It has been widely debated whether
tunneling measurements can see the VHS\cite{VHS2,VHS3}. The results
are quite revealing and show that over the entire doping range
studied, the feature conventionally identified as the VHS (orange
dotted line in Fig.~4) is actually a gap feature\cite{MkSCR,Jarrell}
which can be traced to the bottom of the upper magnetic band.  This
explains why no VHS is found in normal state tunneling near optimal
doping when the superconductivity is turned off.  The calculation also
provides insight into the transition to a large-gap insulator at half
filling.  The AF pseudogap is essentially given by the separation
between the VHS and the bottom of the upper magnetic band, features A
and B in Figs.~2(b),~2(c).  From Fig.~2(a), this can be seen to be
approximately 0.3~eV at $x=0.08$, and the route to a 2~eV gap at half
filling is clear.

Instead, we find the true VHS feature at higher energies.  The doping
evolution is consistent with that found in overdoped Bi2201
cuprate\cite{VHS3} where the VHS is clearly seen close to the Fermi
level. The anticorrelation of the VHS peak with superconductivity
could arise because the VHS can drive a competing ferromagnetic
instability.\cite{TallStorey,Bob_sc}

It should be noted that the present model is not consistent
with the conventional picture of peak-dip-hump\cite{pdh}. By a relatively small
change of parameters we have found a second solution, much closer to
this conventional form.  While this model reproduced many of the
features of the STM data, it was ultimately less satisfactory than the
model described in the text.  In Appendix C, we briefly review this
peak-dip-hump model.  Among other problems noted in Appendix C, the
peak-dip-hump model is unable to explain the transition to a
large-gap insulator at half filling, producing a magnetic gap of only
$\sim$100~meV even at the lowest doping.

While cuprates have traditionally been treated in strong coupling
formalism, recent investigations have shown that intermediate coupling
models, such as the present QP-GW model, can capture many salient
features of cuprate physics, including the doping dependence of
dispersion and optical
properties\cite{markiewater,comanac,DMFT2,tanmoyop}. In particular,
our QP-GW self-energy successfully describes the dressing of
low-energy quasiparticles by spin and charge
fluctuations,\cite{markiewater} the high-energy kink (HEK) seen in
ARPES,\cite{markiewater} the residual Mott bands in the optical
spectra,\cite{tanmoyop} gossamer features,\cite{tanmoyz} anomalous
spectral weight transfer with doping\cite{ASWT}, and the magnetic
resonance in neutron scattering.\cite{ArunSNS} In fact, our
intermediate coupling model yields a number of characteristic
signatures of strong coupling physics of the Hubbard model, including
suppression of double occupancy\cite{footB5}, the $t-J$-model
dispersions\cite{footB6}, spin wave dispersion\cite{SWZ}, the $1/U$
scaling of the magnetic order,\cite{footB7} and the phenomenon of
anomalous spectral weight transfer (ASWT)\cite{footB8,ASWT0,ASWT}.
Interestingly, a recent variational calculation finds a smooth
evolution from a SDW to a Mott gap with increasing $U$, with no
intervening phase transition or spin liquid phase in the cuprate
parameter range.\cite{TBPS}

\section{Summary and conclusions}

We have carried out a realistic, multiband modeling of STS spectra in
Bi2212 over a wide doping range where effects of competing pseudogap
and superconducting gaps are incorporated, and strong electron
correlations are treated by introducing self-energy corrections in the
one-particle propagators. The theoretical spectra capture many of the
salient features of the corresponding experimental spectra and their
doping dependencies, including the two-gap behavior at low energies,
electron-hole asymmetry of the coherence peaks, and the presence of a
prominent hump feature at high energies. Our analysis yields insight
into the complex manner in which the VHS manifests itself in the STS
spectra.

{\bf Acknowledgments} It is a pleasure to acknowledge technical
assistance of Mr. Ray Wang. This work is supported by the US
Department of Energy, Office of Science, Basic Energy Sciences
contract DE-FG02-07ER46352, and benefited from the allocation of
supercomputer time at NERSC, Northeastern University's Advanced
Scientific Computation Center (ASCC) and the resources of Institute of
Advanced Computing, Tampere.  I.S would like to thank Ulla Tuominen
Foundation for financial support.

\appendix

\section{Details of interaction terms in the Hamiltonian}

Here we provide details concerning the superconducting and
antiferromagnetic terms in Eq.~\ref{hamiltonian}.  The mean field
superconducting coupling between electrons and holes is of the form
\begin{equation}
  \label{eq:gapgen}
  \Delta_{\alpha \beta} = \sum_{a b} U_{\alpha  \beta a b} \langle c_{a
  \downarrow}c_{b \uparrow} \rangle. 
\end{equation}
We assume a $d$-wave form for the gap as is appropriate in the
cuprates.\cite{NLMB} The interaction $U_{\alpha \beta a b}$ is not
known, and hence we follow the common practice of introducing a gap
parameter, which gives the correct gap width and symmetry
\cite{Flatte}.  The gap values of $\vert\Delta\vert = 25-45 meV$,
plotted as violet squares in Fig,~4, are chosen to model the
experimental spectra\cite{McElroy,footB1}.  It is interesting to note
that the gap calculated from a spin fluctuation mechanism\cite{Bob_sc}
(red circles in Fig.~4) has a very similar doping dependece, andeed a
very similar magnitude to the smaller gap feature.

Based on the QP-GW results, we incorporate magnetic interactions into
our model via an onsite interaction term
\begin{equation}
  \label{eq:inter}
  H^{C} = U_{\alpha \alpha} n_{\alpha \uparrow}n_{\alpha \downarrow}
= U_{\alpha \alpha} c^{\dagger}_{\alpha \uparrow}c_{\alpha \uparrow}
c^{\dagger}_{\alpha \downarrow}c_{\alpha \downarrow}.
\end{equation}
At mean field level, this reduces to a magnetic interaction
$$-\frac{1}{2}U_{\alpha \alpha} 
\langle m_{\alpha} \rangle \left( c^{\dagger}_{\alpha
    \uparrow}c_{\alpha \uparrow}- c^{\dagger}_{\alpha
    \downarrow}c_{\alpha \downarrow} \right),$$
where $m_{\alpha} = n_{\alpha \uparrow} - n_{\alpha \downarrow}.$
In the present calculation, we assume a doping independent $U=7.5~eV$
and the $<m_{\alpha}>$ are found self-consistently from a mean field
calculation.\cite{Hsin,MBRIXS,footB1}

\section{Details of Self-energy corrections}

Details of the three contributions to the self-energy of
Eq.~\ref{eq:totalself} are as follows.  In the phonon contribution
$\Sigma^L$, the phonons are approximated by a Debye
spectrum\cite{Pickett, Cohen}.  The self-energy can then be written
as a convolution of the Debye spectrum with a constant density of
states for the electrons as:\cite{nieminenPRB}
\begin{equation}
\Sigma^{L}(\varepsilon) =
 -\frac{A}{\pi} \left( (2z+i\pi) + \left(z^2 - 1\right)
 \ln{\left(\frac{z - 1} {z + 1} \right)}
 \right),
\label{sigmaeinstein}
\end{equation}
where $z=(\varepsilon + i \eta)/(\hbar \Omega_d)$,
$A = \frac{3 \hbar}{4 \Omega_d} \Gamma^2 \rho,$ and $\eta$ is a
convergence parameter.  We assume the self-energy for simplicity to be
of a diagonal form, and apply this to Cu-$d_{x^2-y^2}$ orbitals using
$\hbar \Omega_d = 80meV$ and $A = 60meV.$ When combined with
$\Sigma^H$, these values yield a reasonable description of the
peak-dip-hump structure and the smoothness of the spectrum.

For the electron-electron self-energy, $\Sigma^H$, we utilize a simple
Fermi-liquid type self-energy\cite{NLMB}, except for the antibonding
$CuO_2$ band, nearest the Fermi level, for which we introduce a QP-GW
self-energy,\cite{tanmoyop}
\begin{eqnarray}
  \label{eq:selfH}
  \Sigma^{H}(k,\omega) &=& \sum_{q} \int \frac{d \omega '}{2 \pi}
\Gamma G(k-q, \omega + \omega')W(q,\omega'),
\end{eqnarray}
where $W = (U^2/2)Im(3 \chi_s + \chi_c)$ and $U$ is appropriate to the
one-band model. The RPA spin/charge susceptibilities are
$\chi_{c/s}(q,\omega) = \chi_{0}(q,\omega)[1 \pm U
\chi_{0}(q,\omega)]^{-1}$, and the bare two-particle correlation
function $\chi_{0}(q,\omega)$ is the convolution of the green's
function.  Self-consistency in $\Sigma$ is achieved by calculating an
average renormalization factor $Z=(1-d\Sigma/d\omega)^{-1}$ with the
QP-GW method.\cite{markiewater,basak,tanmoyz,tanmoyop}.  The vertex
correction is $\Gamma=1/Z$.

For electron doped cuprates, the competing order is $(\pi ,\pi )$
antiferromagnetism, while the origin of the pseudogap is unclear in
the hole-doped cuprates.  We have shown\cite{tanmoy2gap} that the
tunneling spectra are insensitive to the exact nature of the
pseudogap, as long as it represents a density wave order [charge,
spin, or flux phase] at $(\pi ,\pi )$, so here we model the pseudogap
as a $(\pi ,\pi )$ AFM.  Including a $d$-wave SC gap below $T_c$, the
green's function becomes a $4\times4-$tensor.  The QP-GW self-energies
are calculated in a $(\vert k \rangle,\vert k+Q \rangle)$ basis, then
transformed via a unitary transformation to a real-space basis
involving the spin up and spin down sublattices.  In this transformed
self-energy matrix, the off diagonal terms are found to be small and
of varying sign, and are neglected.
The corresponding matrix elements for the spin up and spin down holes
are obtained by using the relations derived in
Ref. \onlinecite{nieminenPRB}.

Finally, an impurity self energy $\Sigma^{imp}(\varepsilon)=-i5.0meV$
is used in the SC phase for all the orbitals, which produces a
reasonable smearing in the low-energy region. For the normal state, a
larger value of $\Sigma^{imp}(\varepsilon)=-i 20.0meV$ is used to
account for enhanced normal state broadening near the Fermi energy.

\section{Peak-Dip-Hump Model}

The calculations described in the main text involve a moderate band
renormalization by correlation effects, corresponding to a
renormalization factor $Z=0.5$, similar to that found in LSCO and
NCCO.  However, in Bi2212 ARPES experiments suggested a much narrower
coherent band, with $Z\sim$0.28.\cite{Arun3} In this Appendix we
briefly describe the results of a model calculation with this smaller
$Z$ value, which also reproduces many of the features
of the data of Figs. 1-4.

This small-$Z$ model reproduces a form of peak-dip-hump scenario.  The
strong renormalization means that the spectral weight below the Fermi
level is split into a broad, weakly renormalized incoherent band
(hump) and a narrow, strongly renormalized coherent band (peak).  In
this model, the broad feature at high energies, with the dispersion in
Fig.~2, is the VHS of the incoherent band, while the coherent band has
a VHS at much lower energies, which varies with doping exactly as the
VHS derived from the ARPES dispersion\cite{MikeN,Arun3}, orange-dotted
line in Fig. 4.  When an AF pseudogap and d-wave superconductivity are
included, this model also reproduces the two-gap physics of Fig.~4.
The dip between these two VHS features closely follows the measured
position of the magnetic resonance peak.\cite{Zas}.

There are several experimental features that this small-Z model does
not successfully reproduce.  For instance, the coherent VHS peak is
always found to be more intense than the incoherent VHS, contrary to
experiment, Fig.~3.  This is not unexpected, as the QP-GW model is
known to underestimate the incoherent spectral weight.  However, there
are additional problems, involving only the low energy, coherent
spectral weight.  Most importantly, there is a true, coherent VHS near
the Fermi level, which should clearly appear as a low energy peak when
the AF and superconducting orders are turned off, in sharp
contradiction to the normal state results plotted in Fig.~1.  A more
subtle feature involves the electron-hole asymmetry of the larger of
the two-gap features as a function of doping.  In this doping range,
the coherent VHS always lies below the Fermi level, so the negative
energy peak should always be more intense, and indeed the peak
asymmetry should increase with underdoping, as the VHS moves further
from the Fermi level.  However, at very low dopings, the experimental
asymmetry reverses sign, Fig.~3(b).  Indeed, this can be understood
from the model calculations of Fig.~2(a).  Feature B, the lower edge
of the UMB, actually crosses the Fermi level, lying above it at low
dopings.  With this crossing, the theoretical asymetry reverses, in
good agreement with experiment, Fig.~3(b).  Finally, while the large-Z
model clearly extrapolates to a large-gap insulator at half filling,
the small-gap model does not. The magnetic gap is directly reflected
in the two gap physics near the Fermi level.  Hence, even at the
lowest doping it remains of order $\sim$100~meV, and the manner in
which a $\sim$2~eV gap arises at half filling remains a puzzle.  It is
for these reasons that we prefer the larger-$Z$ model.  The apparently
accidental pinning of the bottom of the UMB near the Fermi level may
actually be a signature of a cooperative interaction between the AF
and superconducting order, thereby stabilizing both phases.


\begin{thebibliography}{99}
\bibitem{footB4}The one-band QP-GW model for the non-superconducting state
  has only one free parameter, which is the value of the Hubbard $U$
  at half filling, since the hopping parameters are taken
  from LDA and the doping dependence of $U$ is calculated from
  screening by long-range Coulomb interactions.
\bibitem{footB10}This assumption has however been
  questioned\cite{McElroy,Nunner}.
\bibitem{foot11}Spin-flip terms are neglected in the present
calculations.
\bibitem{abfoot1} The doping dependence of the band structure is
  treated within the framework of the rigid band model
  [Ref. \onlinecite{new1}], which is expected to be a good
  approximation for doping away from the CuO$_2$ planes. It will be
  interesting to further examine doping effects by using first
  principles approaches \cite{new2}.
\bibitem{footB1}The gap values are $\Delta$ = $(25,35,35,45,45)$~meV
  and the magnetization values $m$ are = $(0,0.04,0.08,0.12,0.16)$,
  for $x$ = (0.19,0.16,0.12,0.10,0.08).
\bibitem{footB2}Comparing Figs. 1 and 2 at high positive energies, the
  experimental spectra are seen to be quite different. Such an effect
  is neither expected nor seen in the theoretical spectra. The origin
  of this sample dependence of the spectra is not clear.
\bibitem{footB3}When the imaginary part of the self-energy is
  non-zero, the Green's function no longer possesses real poles, and
  strictly speaking, we cannot think in terms of the usual energy
  bands. It can be useful however to think in terms of complex energy
  bands defined by the complex energy solutions of $G(E,k)^{-1}=0$ as
  a function of k. See, Ref. \onlinecite{bansil12}
\bibitem{footab} We should keep in mind that the tunneling matrix
  element plays a substantial role in modulating the LDOS to yield the
  STS spectra. Such matrix element effects are important in
  ARPES\cite{arpesab}, inelastic light
  scattering\cite{MBRIXS,new2,comptonab} and other
  spectroscopies\cite{positronab}.
\bibitem{foot2}These are matrix elements of the Green function between
  majority/minority spin electrons at a chosen site and the holes with
  the opposite spin at a neighbouring site ($F_{\alpha \beta maj} =
  \langle e_{\alpha\sigma(maj)} \vert \mathbf{G} \vert h_{\beta
    -\sigma(maj)} \rangle$ and $F_{\alpha \beta min} = \langle
  e_{\alpha \sigma(min)} \vert \mathbf{G} \vert h_{\beta -\sigma(min)}
  \rangle$).  Their imaginary part measures the pairing amplitudes of
  electrons with opposite spins.
\bibitem{foot5} While we have modeled the pseudogap as due to AFM
  order, any other competing $(\pi ,\pi )$ density-wave order would
  work equally well\cite{tanmoy2gap}.
\bibitem{footB7}A characteristic feature of Mott physics is that as
  $U$ increases the Mott gap grows but the magnetic transition
  temperature ultimately decreases.  This follows in the present
  formalism: even though the gap grows, the energy difference between
  any two magnetically ordered phases scales with $J\sim 1/U$.
\bibitem{footB8}Doping a Mott insulator leads to anomalous spectral
  weight transfer (ASWT): removing one electron from a Mott insulator
  leads to two empty states at low energies (in the lower Hubbard
  band), since there is no longer a penalty $U$ to be paid when the
  spin-reversed electron occupies the state\cite{ASWT0}.  The QP-GW
  model captures most of this effect, due to the real collapse of the
  magnetic gap at a quantum critical point.\cite{ASWT} QP-GW
  underestimates this ASWT in the very low doping regime, although
  here strong disorder may mask this effect, and render it
  unobservable.
\end{thebibliography}
\end{document}